\documentclass[preprint]{elsarticle2}

\usepackage{epic,epsfig,axodraw}
\usepackage{subfigure}
\usepackage{times}

\def\be{\begin{equation}}
\def\ee{\end{equation}}
\def\bea{\begin{eqnarray}}
\def\eea{\end{eqnarray}}
\def\ba{\begin{array}}
\def\ea{\end{array}}
\def\ghost#1{}

\def\beq{\begin{equation}}
\def\eeq{\end{equation}}
\def\bey{\begin{eqnarray}}
\def\eey{\end{eqnarray}}

\def\lsim{\mathrel{\raise.3ex\hbox{$<$\kern-.75em\lower1ex\hbox{$\sim$}}}}
\def\gsim{\mathrel{\raise.3ex\hbox{$>$\kern-.75em\lower1ex\hbox{$\sim$}}}}

\begin{document}
\baselineskip=16.5pt

\title{
$U(1)_A\,$ symmetry in two-doublet models, \vspace{3mm}\\
$\ U$ \,bosons \,or \,light \,scalars,
\,and $\,\psi\,$ and $\,\Upsilon\,$ decays\vspace{3mm}\\}
\author{Pierre Fayet \vspace{-2mm}\\}
\address{Laboratoire de Physique Th\'eorique de l'ENS, UMR 8549 CNRS, 24 rue Lhomond, 75231 Paris Cedex 05, France
\vspace{4mm}\\
{March 20, 2009}
\vspace{-1mm}
\\}

\begin{abstract}

$\psi\,$ and $\Upsilon$ decays
may be used to search for light neutral spin-1 or spin-0 bosons 
associated with a broken extra-$U(1)$ symmetry, local or global, acting axially on quarks and leptons,
as may be present in supersymmetric theories with a $\lambda \ H_1 H_2\,S\,$ superpotential term. 
Recent data on $\ \Upsilon\to \gamma\,+$ {\em invisible neutral\ } constrain
an axial, pseudoscalar or scalar coupling to $\,b\,$ to  $\,f_{bA}<4\ 10^{-7}\ m_U$ (MeV)$/\sqrt{B_{\hbox{\scriptsize inv}}}\,$,
\,$f_{bP}< 4\ 10^{-3}/\sqrt{B_{\hbox{\scriptsize inv}}}\ $ or $\,f_{bS}< 6\ 10^{-3}/\sqrt{B_{\hbox{\scriptsize inv}}}\,$, \,respectively.
This also constrains, from universality properties, couplings to electrons to 
$\ f_{eA}< \,4\ 10^{-7}\, m_U$(MeV)$/\sqrt{B_{\hbox{\scriptsize inv}}}\,$, 
$\ f_{eP}< \,4\ 10^{-7}/\sqrt{B_{\hbox{\scriptsize inv}}}\ $ \,or $\ f_{eS}< \,6\ 10^{-7}/\sqrt{B_{\hbox{\scriptsize inv}}}\ $.
\vspace{1.2mm}\\
\indent
The pseudoscalar $\,a\,$ (possibly traded for a light gauge boson, or scalar particle)
should then be, for invisible decays of the new boson,
\,for $\,>96\,\%\,$ singlet and $\,<4\,\%\,$ doublet, for $\,\tan\beta>1$.
Or, more generally, $\ <\,4\,\%\,/(\tan^2\beta \,B_{\hbox{\footnotesize inv}}$)\, doublet, which 
implies a very small rate for the corresponding $\psi\,$ decay,
$\ B\, (\psi\to\gamma\, +$ {\em neutral\,)} $ B_{\hbox{\footnotesize inv}}\ \lsim \,10^{-6}/\tan^4\beta\,$.
\,Similar results are obtained for new spin-1 or spin-0 neutral bosons 
decaying into $\ \mu^+\mu^-$.

\end{abstract}

\maketitle

\vspace{3mm}

\noindent
PACS numbers: 12.60.Cn \ 12.60.Fr \ 12.60.Jv \ 13.20.Gd \ 14.70.Pw \ 14.80.Mz
\hfill LPTENS-08/70

\vspace{3mm}

\section{$\,U(1)$ \,symmetries in two-doublet models}

Particle physics theories involving {\,\em two\,} Englert-Brout-Higgs doublets, now denoted as  
$\,(h_1^\circ,\ h_1^-)$ and  $\,(h_2^+,\ h_2^\circ)$, allow for a possible
$\,U(1)$ symmetry acting as 

\vspace{-5mm}
\be
\label{u10}
h_1\ \to \ e^{i\alpha}\ h_1\ ,\ \ \ \ h_2\ \to \ e^{i\alpha}\ h_2\ \ ,
\ee
constraining 
both their interaction potential and Yukawa couplings to fermions
\cite{qinv} \footnote{\small The allowed quartic interactions in $\,V(h_1,h_2)\,$ are 
$\,(h_1^\dagger h_1)^2,\,(h_2^\dagger h_2)^2,
\ (h_1^\dagger h_1)\,(h_2^\dagger h_2)$ \,and $\,|h_1 h_2|^2\,$ 
or equivalently $\ |\,h_1^\dagger  h_2\,|^2=
(h_1^\dagger h_1)\,(h_2^\dagger h_2)\, - \,|h_1 h_2|^2$\,.
\ Within supersymmetry they appear as electroweak gauge interactions, with \,\cite{fayetsinglet} 
$$
V_{\,\rm quartic}\ \ =\ \ \frac{g^2+g'^2}{8}\ \ (h_1^\dagger h_1-h_2^\dagger h_2)^2\,+\,\frac{g^2}{2}\ \ |\,h_1^\dagger h_2|^2\ \ .
$$
}.
This occurs naturally within supersymmetric extensions of the standard model, which require two doublet 
superfields $\,H_1\,$ and $\,H_2\,$ responsible for the electroweak breaking and the generation of quark and lepton masses \cite{fayetsinglet, susy1}. Such a transformation
may also be used as a possible way to rotate away $\,CP$-violating effects in QCD \cite{pq}.

\vspace{1mm}
This $\,U(1)\,$ symmetry, broken through $\,<\!h_i^{\,\circ}\!>\ =v_i/\sqrt 2\,$
\,with $\,\tan\beta=v_2/v_1\,$, would lead, after the Goldstone combination
$\ \hbox{Im}\, (\cos\beta \, h_1^{\,\circ}\!-\sin\beta\, h_2^{\,\circ})$
gets eliminated by the $Z$, to a quasi-massless ``axion'' field

\vspace{-4mm}

\be
\label{A}
A\ =\  \sqrt 2\ \ \hbox{Im}\, (\,\sin\beta \, h_1^{\,\circ}+\,\cos\beta\, h_2^{\,\circ}\,)\ \ ,
\ee

\vspace{-1mm}

\noindent
if it were not broken explicitly 
as in \cite{fayetsinglet}\,\footnote{\small We used $\,i\,\varphi"=h_1$, \,$\,i\,\varphi'=h_2^{\,c}\,$
\,(described by left-handed and right-handed doublet superfields respectively), 
with $\,\varphi"\to e^{i\alpha}\,\varphi",\ \varphi'\to e^{-\,i\alpha}\,\varphi'$, 
under a $\,U$ (or $\,Q=RU$) transformation. The mixing angle $\,\beta\,$ was called $\,\delta\,$, 
with $\tan\delta=v'/v"$ instead of the present $\tan\beta=v_2/v_1$. The symmetry (\ref{u10})
acting on the two Higgs doublets, called $\,Q\,$ in the 2-doublet pre-SUSY model of \cite{qinv}, 
was extended to supersymmetry in~\cite{fayetsinglet} according to 

\vspace{-2mm}
$$
H_1 \ \stackrel{Q}{\to} \ e^{i\alpha}\ H_1(x,\theta\,e^{-i\alpha})\,,\ \ \ 
H_2 \ \stackrel{Q}{\to} \ e^{i\alpha}\ H_2(x,\theta\,e^{-i\alpha})\,,\ \ \ \hbox{and} \ \ \ 
V(x,\theta,\bar\theta)\ \stackrel{Q}{\to} \ V(x,\theta\,e^{-i\alpha},\bar\theta\,e^{i\alpha})\,
$$
for gauge superfields.
This original $Q$ symmetry was an ``$R$-type'' symmetry,
allowing for a direct $\,\mu\,H_1 H_2$ mass term but which did not survive electroweak breaking. 
It was transformed \cite{fayetsinglet} into the true $R$-symmetry (the progenitor of $R$-parity $R_p=(-1)^R$) \,acting as

\vspace{-2mm}
$$
H_1\ \stackrel{R}{\to} \  H_1(x,\theta\,e^{-i\alpha})\,,\ \ \
H_2\ \stackrel{R}{\to} \ H_2(x,\theta\,e^{-i\alpha}), \ \ \ \hbox{and} \ \ \ V(x,\theta,\bar\theta)\
\stackrel{R}{\to} \ V(x,\theta\,e^{-i\alpha},\bar\theta\,e^{i\alpha})
$$
through a global $U(1)\,$ transformation commuting with supersymmetry,
$\ U^{-1}=R\,Q^{-1}\,$  with $\,U$ acting as in (\ref{u2}), \,now called $\,U(1)_A$.}. 
This explicit breaking through $f(S)$ superpotential terms provides a mass, 
proportional to the $\lambda$ parameter of the $\,\lambda\,H_1 H_2\,S\,$ superpotential coupling 
with the singlet $S$ introduced in \cite{fayetsinglet}, for the ``axion'' field A (\ref{A}) 
that would otherwise remain quasi massless.
This extra-$U(1)$, \,which acts axially on quarks and leptons and will be referred to as $U(1)_A$,
\,may also be taken (in the absence of $f(S)$ and $\,\mu\,H_1 H_2\,$ superpotential terms that would break it explicitly) 
\,as a {\it \,gauged symmetry}, 
leading to the USSM \cite{susy1}.
It is very much the same as the $\,U(1)$ considered in \cite{pq}, 
excepted that anomalies should in principle be cancelled if this $\,U(1)\,$ is to be gauged.
The pseudoscalar Goldstone boson eaten away in \cite{susy1}
to give a mass to the new neutral gauge boson of $\,SU(3)\!\times\! SU(2)\!\times \!U(1)\, \times$ extra-$U(1)$ 
is very similar to the axion found later in \cite{wilczekaxion,weinberg}. When the extra-$U(1)$ is gauged, the 
new gauge boson acquires its mass by eliminating a would-be~\,``axion''.

\vspace{1mm}

This neutral gauge boson, referred to as $\,U$ (also often called $Z'$), \,did not show up in neutral current phenomenology 
nor direct searches at particle colliders. It can be made much heavier 
than the $\,W^\pm$ and $Z$, \,say of $\sim$ TeV scale, as the singlet described by $\,S\,$
can acquire a large v.e.v.,
making the $U$ contribution to neutral current effects 
sufficiently small~\cite{fayetu0}.
The mass  $\,m_U=g" F_U$, on the other hand, may be small if the extra-$U(1)$ gauge coupling $g"$ is small or very small.
However, even in the case of a very small gauge coupling the $U$ could still conserve sizeable interactions, as it would in fact behave very much 
as the eaten-away Goldstone boson \,$A\,$ in (\ref{A}), in the absence of a singlet v.e.v.;
\,or in a more general way as a doublet-singlet combination $\,a$,
which would make it much harder to detect \cite{fayetu0,fayetu}.
This was used long ago to discuss the production of
light spin-1 $U$ bosons or of their effectively-equivalent spin-0 pseudoscalars $\,a$
in the radiative decays \ $\psi\,$ and 
$\,\Upsilon\,\to\,\gamma\ \,U/a\,$ \cite{fayetu,fayetmezard}.

\vspace{1mm}
A very light $\,U\,$ does not decouple in the limit $\,g"\to 0$, \,but gets produced and interacts very much 
as the eaten-away axionlike pseudoscalar $a$ \cite{fayetu}. 
In the absence of a singlet v.e.v. this one is a mixing of $\,h_1^{\,\circ}\,$ and $\,h_2^{\,\circ}\,$ 
as defined by $A$ in (\ref{A}), and would be produced as a standard axion, 
a possibility that turned out to be excluded.
When the extra-$U(1)$ is broken not only by $\,<\!h_1^{\,\circ}\!>\,$ and $\,<\!h_2^{\,\circ}\!>\,$ but also 
by a large singlet v.e.v. $<\!s\!>\,$,
\,at a scale possibly significantly larger than the electroweak scale,
the spin-1 $U$  boson is produced and interacts as the (eaten-away) axionlike pseudoscalar $\,a$,
now given by the doublet-singlet combination

\vspace{-5mm}
\be
\label{a}
\hbox{\em pseudoscalar}\ \ a\ \,=\,\ \cos\zeta \ \ \left(\,\sqrt 2\ \ \hbox{Im}\ (\sin\beta \ h_1^{\,\circ}\,+\,\cos\beta\ h_2^{\,\circ}\,\right)
+\ \sin\zeta \ \ (\,\sqrt 2\ \ \hbox{Im}\ s\,)\ \ .
\ee

\vspace{-1mm}
\noindent
The pseudoscalar $A$ (with the same expression (\ref{A}) as for the standard axion, 
or $A$ of the MSSM) 
mixes with the singlet~$s$,
\linebreak
\vspace{-9.5mm}

\pagebreak

\noindent
uncoupled to quarks and leptons and to electroweak gauge bosons. 
The resulting combination $a$ thus interacts essentially through its doublet component $\,A\,\cos\zeta$, 
\,proportionally to the
{\em \,invisibility parameter} $\,r=\cos\zeta\,$.
\,The branching ratios for $\,\psi\,$ or $\,\Upsilon\to\gamma+U/a$ are essentially the same as for a standard axion 
 $\,A$ \cite{wilczekaxion} but multiplied by 
$\,r^2=\cos^2\zeta\,$ \cite{fayetu,fayetmezard,melange}.
If $\, <\!s\!>\,$ is large, the $U(1)$  is broken 
{\em ``at a large scale''}, 
$\,r=\cos\zeta\,$ is small, and the pseudocalar $\,a\,$ 
(or associated $\,U$ boson in case of a local $\,U(1)$ symmetry) 
\,becomes largely ``invisible''. 
This mechanism can be used as well for a spin-1 $U$ boson or spin-0
axion or axionlike pseudoscalar $a$ (or even also scalar), \,then mostly an electroweak singlet as proposed in \cite{fayetu0}, 
according to what was called later the ``invisible axion'' mechanism.

\vspace{1mm}

If the extra-$U(1)$ is only global (and possibly anomalous)
and broken {\,\it ``almost spontaneously''\,} but with small additional explicit-breaking terms,
the would-be Goldstone or quasi Goldstone boson $a$ acquires small mass terms, with its production 
rates still given by the same formulas, proportionally to $\,r^2=\cos^2\zeta$.
\,This applies, in particular in the N/nMSSM, when the $\,U(1)$ symmetry considered
(which may be a $\,U(1)_A\,$ commuting with supersymmetry, or an $R$-symmetry not commuting with it)
\,is explicitly broken in this way through {\it \,small\,} superpotential couplings 
\,(such as $\,\frac{\kappa}{3}\, S^3$, $\,\frac{\mu_S}{2}\, S^2,\ \sigma S$, \,or $\,\mu\,H_1 H_2$) 
\,and/or {\it \,small\,} soft supersymmetry-breaking terms, ultimately responsible for a small mass 
for the pseudoscalar $a$ associated with this ``almost spontaneous'' breaking of the global $\,U(1)$.

\vspace{1mm}

We shall be especially interested in spin-1 $U$ bosons associated with a local extra-$U(1)$ symmetry 
as in the USSM~\cite{susy1}, which may decay into $\,\nu\bar\nu\,$\ or  $\,e^+e^-,\,$ ... \,, depending on their mass \cite{fayetu} (or even have dominant decays into light dark matter particles \cite{ldm,fayetldmu}), 
light pseudoscalars being discussed in \cite{dg}.
More generally we shall obtain, from $\,\Upsilon\,$ decays, new constraints on the pseudovector, pseudoscalar or scalar 
couplings of the new boson to the $b$ quark. They have important implications for the decay
$\,\psi\to \gamma$ + {\it \,invisible neutral}, \,which should be very small, as well as for the couplings of the new spin-1 or spin-0 boson to {\it charged leptons}. Finally, we also discuss new constraints obtained from searches for
a neutral boson decaying into $\,\mu^+\mu^-$, \,and their implications for $\,\psi\,$ decays and new boson couplings.

\vspace{1mm}

\section{\boldmath Extra \,$U(1)_A\,$ and \,extra \,singlet \,from \,supersymmetric \,theories}

Having two Higgs doublets instead of a single one as in the standard model 
allows for the possibility of 
``rotating'' them independently, thanks, in addition to the weak hypercharge $\,U(1)$, 
\,to the extra-$\,U(1)\,$  symmetry acting as  \cite{qinv}
\be
\label{u1}
h_1=\ \left(\,\ba{c} h_1^{\,\circ} \vspace{2mm} \\ h_1^- \ea\ \right)
 \ \ \stackrel{U}{\longrightarrow}\ \ e^{i\,\alpha}\ \ h_1\ \ ,\ \ \ \ 
h_2=\ \left(\,\ba{c} h_2^+ \vspace{2mm} \\ h_2^{\,\circ} \ea\ \right) \ \ \stackrel{U}{\longrightarrow}\ \ e^{i\,\alpha}\ \ h_2\ \ ,
\ee
embedded within supersymmetric models according to \cite{fayetsinglet} 
\be
\label{u2}
H_1 \ \ \stackrel{U}{\longrightarrow}\ \ e^{i\,\alpha}\ \ H_1\ \ ,\ \ \ \ 
H_2\ \ \stackrel{U}{\longrightarrow}\ \ e^{i\,\alpha}\ \ H_2\ \ ,
\ee
and broken through 
$\ <h_1^{\,\circ}\!>\ $ and $\ <h_2^{\,\circ}\!>\  $.
\,The $\,\mu\,$ parameter of the $\,\mu\,H_1 H_2\,$ superpotential term, not invariant under this extra $\,U(1)$
\footnote{\small This  $\,U(1)_A\,$ symmetry is broken explicitly in the MSSM through $\,\mu\,H_1 H_2\,$ 
and by a soft susy-breaking term proportional to \,Re\,$\,h_1h_2$, \,allowing the pseudoscalar 
$A$ in (\ref{A}) to acquire a mass. 
\vspace{.5mm}\\
\phantom{a}\hspace{2mm}
Note that there is no specific hierarchy problem associated with the size of the supersymmetric $\mu$
parameter, \,which may be kept small (compared to large masses like $m_{GUT}\,$ or $m_{\rm Planck}\,$)
\,by means of this (broken) extra-$U(1)_A$ symmetry. Or also by a continuous 
(broken) $U(1)_R$-symmetry, so that 
$\mu\,$  may be naturally of the same order as susy-breaking parameters, most notably gaugino masses $m_{1/2}$~\cite{cfg}
which break explicitly this continuous $U(1)_R$.}
(nor under the continuous $R$-symmetry),
\,was then promoted into a dynamical variable 
$\,\mu(x,\theta)$, \,with $\,\mu\,H_1 H_2\,$  replaced in \cite{fayetsinglet} 
by a trilinear coupling $\,\lambda\ H_1 H_2\,S\,$ with the extra singlet $S$, \,transforming under 
$\,U\,$ 
as
\be
\label{UsurS}
S\ \ \stackrel{U}{\longrightarrow} \ \ e^{-\,2\,i\,\alpha}\, S\ \ .
\ee
This $\,U(1)\,$ symmetry acts axially on quark and lepton superfields 
according to  \cite{susy1}
\footnote{\small This $\,U(1)_A$ symmetry was initially also introduced and gauged, possibly with a very small coupling $\,g"$, 
\,in view of generating spontaneously (universal) squark and slepton mass$^2$ terms $\,m_\circ^2\,$ through $\,<D>\ $ contributions 
of the extra-$U(1)_A$, 
\,leading in the simplest case \cite{sugra,fayetbrisure} \,to 

\vspace{-6mm}
$$
\,m_\circ^2\,(\tilde q, \tilde l)\ =\ \frac{g"}{4}\ <\!D\!>_A\ \ .
$$
This also illustrates the connection between a very weakly coupled $\,U$, with $\,g"$ very small, and supersymmetry broken 
``at a large scale'' with a very weakly coupled goldstino/gravitino \cite{sugra}, and how soft susy-breaking terms may be generated spontaneously, when the susy-breaking scale gets very large 
so that the goldstino decouples. \vspace{1mm}}
\be
\label{ulq}
(Q,\bar U,\bar D;\,L,\bar E)\ \ \stackrel{U}{\longrightarrow}\ \ 
e^{-\,i\,\frac{\alpha}{2}}\ (Q,\bar U,\bar D;\,L,\bar E)\ \ ,
\ee
and may be referred to as $\,U(1)_A$ \footnote{\small More generally the $U(1)_A$ symmetry considered may be replaced by a $\,U(1)\,$ 
associated with a linear combination of the $\,U(1)_A$ generator $F_A$ with $\ \alpha B+\beta_i L_i+\gamma Y$.\vspace{1mm}}. 
The superpotential $\,{\cal W}\,$  (with the $\,\mu\,H_1H_2$ term of the MSSM absorbed within the trilinear $\,\lambda \, H_1 H_2\,S\,$
superpotential) \,may be written in a general way (omitting family indices for simplicity) as 
\be
\label{trilq}
\ba{c}
\\[-3mm]
{\cal W}  \ \ = \ \ \underbrace{\,\lambda_e \ H_1 \,.\,\bar E \,L \ +\ 
\lambda_d \ H_1\,. \,\bar D \,Q \ -\  
\lambda_u \  H_2 \,.\,\bar U \,Q\,}_{\stackrel{}{\hbox{\normalsize ${\cal W}_{\,lq}$}}}\ \ +\ \ \lambda \ H_1 H_2\,S\ \ +\ \ 
\underbrace{\,\frac{\kappa}{3}\ \,S^3\ +\ \frac{\mu_S}{2}\ \,S^2\ +\ \sigma\,S\,}_{\hbox{\normalsize$f(S)$}}\ \ .
\vspace{-1.5mm}\\
\ea
\ee
${\cal W}_{\,lq}$,  \,responsible for quark and lepton masses, and
the $\,\lambda \ H_1 H_2\,S\,$ superpotential coupling with the singlet $S$, are both
invariant under the extra-$U(1)$ symmetry, as well as under the continuous $\,U(1)_R$ symmetry 
that led to $\,R$-parity, $R_p=(-1)^R$ \footnote{\small By ``invariant under $\,U(1)_R$ symmetry'' 
we mean that the superpotential terms ${\cal W}_{\,lq}$  \,and $\,\lambda \,H_1 H_2\,S\,$
transform according to 
$$
{\cal W}\ \ \to \ \ e^{2\,i\,\alpha}\ {\cal W}(x,\theta e^{-i\alpha})\ \ ,
$$ 
so that their $F$-components, proportional to \,Re {\tiny $\int$} $\!{\cal W}\, d^2\theta$,  \,are $R$-invariant.
They are also invariant under any modified $\,U(1)_R$ symmetry 
combining the original $\,U(1)_R$ with $\,U(1)_A$, \,as generated by
$\,R'=R+c\,F_A\,$ (or $\,R+c\,F_A+\alpha B+\beta_i L_i+\gamma Y$) \,in which $\,F_A$ is the generator of $\,U(1)_A$.
\vspace{1mm}
}. 
The terms $\,f(S)\,$, \,which provide in the N/nMSSM\,\footnote{\small In the nMSSM the superpotential (\ref{trilq}) 
is further restricted to 
$$
{\cal W}_{\hbox{\footnotesize\em nMSSM}}\ \,=\,\ {\cal W}_{\,lq}\,+\,\lambda  \  H_1 H_2\,S \,+\,\sigma S\ ,
$$
using the $\,U(1)_R$-symmetry, acting as 
$\,S \stackrel{\hbox{\tiny$R$}}{\to} e^{2i\alpha}\ S(x,\theta\,e^{-i\alpha})\,$,
\,which restricts $f(S)\,$ to the linear $\,\sigma S\,$ term \cite{fayetsinglet}. 
This superpotential already allows for electroweak breaking, even in the absence of supersymmetry-breaking terms.
\vspace{1mm}
}
an explicit breaking of the extra-$U(1)$  
so that there is no quasi-massless ``axion'',
are excluded if this $\,U(1)$ is gauged,
as well as the direct mass term $\,\mu\,H_1 H_2$ \,(that may be dynamically restored
from $\,\lambda \ H_1 H_2\,S\,$
through \,$<\!s\!>\,$). \,Then (\ref{trilq}) reduces to the superpotential of the USSM, 
\be
{\cal W}_{USSM}\  \,=\ \ {\cal W}_{\,lq}\,+\ \lambda \ \,H_1 H_2\,S\ \ ,
\ee
the would-be ``axion'' 
being eliminated when the new gauge boson acquires its mass \cite{susy1} \footnote{\small See e.g.~\cite{kali} for a recent study of neutralino dark matter in the USSM with a heavy $\,U$ boson, and \cite{quevedo} for a discussion of light, very weakly coupled gauge bosons in string compactifications.}.

\vspace{2mm}

\section{\boldmath Axial \,and \,pseudoscalar \,couplings of $\,U\,$ and $\,a\,$ \,to \,quarks \,and\, leptons}

\vspace{1.5mm}

\vbox{
The mass $\,m_U\,=\,g"\, F_U$ \,(with $\,F_U\,$ representative of new symmetry breaking scale)
\,may be naturally small if the extra-$U(1)$ gauge coupling $\,g"$ is small. 
One might think that the $\,U\,$ should then decouple, as the amplitudes for emitting (or absorbing) it,
$\ 
{\cal A}\, (A \to B+U)\,=\, g"\, (\, ...\, ),\,$
seem to vanish with $\,g"$. 
But the longitudinal polarisation vector $\,\epsilon^\mu_{\,L}\,\simeq \,k^\mu_{\,U}/m_U\,$ then becomes singular, so that 
\be
\label{ampllong}
{\cal A}\ (\,A\,\to\, B\, +\, U_{\rm long}\,)\ \ \ \propto\ \ \ g"\ \ \frac{k^\mu_{\,U}}{m_{U}}\ 
<B\,|J_{\mu\,U}|\,A > \ \ \ \ = \ \
\frac{1}{F_U}\ \ k^\mu_{\,U}\ 
<B\,|J_{\mu\,U}|\,A > 
 \ \ 
\ee

\vspace{-7mm}
}

\noindent
has a finite limit.
A very light $U$ with longitudinal polarisation 
couples proportionally to $\,f_{V,A}\,k^\mu_{\,U}/m_U\,$, \,where  $\,k^\mu_{\,U}\,$ acting on an {\it \,axial\,} current $\,\bar f\,\gamma^\mu\gamma_5\,f$
resurrects an effective pseudoscalar coupling to $\,\bar f\,\gamma_5\,f$ with a proportionality 
factor $\,2\,m_f$.  
\,A light $U$ has thus effective
{\it \,pseudoscalar\,} couplings, in particular to quarks and leptons, given in terms of original axial ones by

\vspace{-7mm}
\be
\label{effps}
f_{q,l\ P}\ \ =\ \ f_{q,l\ A}\ \ \frac{2\ m_{q,l}}{m_{U}}\ \ .
\ee
This equivalence theorem ensures that
a light $\,U$ 
with non-vanishing {\it \,axial\,} couplings to fermions
behaves very much as the ``eaten-away''  {\it \,pseudoscalar} $\,a$ \cite{fayetu} 
\footnote{\small The $U(1)$ coupling  $\,g"$\ may be taken as small or very small,
and the $\,U$ related (in part) to the gravitino as it participates in spontaneous supersymmetry breaking through a non-vanishing 
$\,<\!D\!>\ $, \,contributing in particular to the mass$^2$-splittings $\,m_\circ^{\,2}\,$ for
squarks and sleptons \cite{fayetbrisure}.
The spin-1 $\,U$ boson, which has eaten away the axionlike pseudoscalar $a$, is then partly related to the spin-$\frac{1}{2}$ goldstino eaten away by the spin-$\frac{3}{2}$ gravitino, partner of the spin-2 graviton \cite{fayetu0,fayetu}.  Due to this relation with gravity,
$\,g"$ appears as $\,\propto$ $\kappa\,$ and possibly very small.\vspace{1mm}}
\footnote{\small Anomalies associated with the extra-$U(1)$ should
in principle be cancelled (using e.g. mirror fermions or $E_6$-like representations, ...)
if it is to be gauged, and we assume this is realized. 
However, due to the relation with gravity,
with $\,g"$ possibly very small,
the cancellation of anomalies may not be necessary within the low-energy field theory \cite{fayetu0},
and could involve other sectors related to gravity or strings
(see e.g.~\cite{vafaetal} for a related discussion).\vspace{1mm}}.
This is perfectly analogous to what happens 
for a light spin-$\frac{3}{2}\,$ gravitino, whose $\,\pm\,\frac{1}{2}\,$ polarisation states, 
although coupled with gravitational strength $\,\propto\,\kappa$, \,continue to behave very much as
a spin-$\frac{1}{2}\,$ goldstino, 
according to the equivalence theorem of supersymmetry, and with a strength inversely proportional 
to the supersymmetry-breaking scale parameter  \cite{sugra}
\footnote{\small Its effective interactions
are $\ \propto \kappa/m_{3/2}\ $ i.e. inversely proportional to $\,d\,$ or equivalently $\,\Lambda_{\rm ss}^{\,2}$. \,They may or may not be very small, depending on the scale ($\Lambda_{\rm ss}$) at which supersymmetry is spontaneously broken. With
$$
m_{3/2}\ = \ \kappa \,d/\sqrt 6\ = \ \kappa\,F/\sqrt 3\ =\ \sqrt{{\scriptsize \,8\pi\,G_N/3}} \ \ F \ \ ,
$$
 one has 
\,$\,\Lambda_{\rm ss}=\sqrt F=$ $ (\frac{3}{8\,\pi})^{1/4} \sqrt{\,m_{3/2}\,m_{\rm Planck}}$\,. \ A large $d$ \,(i.e.~supersymmetry spontaneously broken ``at a large scale'' $\,\Lambda_{\rm ss}\,$ with a very weakly coupled goldstino component of the gravitino)
\,is then naturally connected with a very small $\,g"$, \,corresponding to a very weakly coupled $\,U$ boson.
\vspace{1mm}
}.

\vspace{3mm}

The couplings of $\,h_1^{\,\circ}$ and $\,h_2^{\,\circ}$ to quarks and leptons are 
$m\,\sqrt 2/(v\, \cos\beta)\,$ and $m\,\sqrt 2/(v\, \sin\beta)$.
The pseudoscalar couplings of $A$ in (\ref{A}) are thus $\,(m/v)\times (\tan \beta=1/x)\,$
for charged leptons and down quarks,
and $\,(m/v)\times (\cot \beta=x)\,$ 
for up quarks, acquiring masses through $\,h_1$ and $\,h_2$, respectively.
With $\,v=2^{-1/4}\ G_F^{-1/2}\!\simeq 246$ GeV  
we get
the pseudoscalar couplings of the standard axion (or $A$ of the MSSM),
$\ 2^{\frac{1}{4}}\,G_F\,\!^\frac{1}{2}$ $m_{q,l}\, 
\times \,(\tan\beta\ $ or $\ \cot\beta)$.
When $\,A\,$ in (\ref{A}) mixes with the imaginary part of 
$s\,$ into expression (\ref{a}) of the pseudoscalar $a$ associated with the extra-$U(1)$ breaking,
we get the following pseudoscalar (or effective pseudoscalar) couplings, now also proportional the invisibility parameter $r=\cos\zeta$,
\be
\label{couplage2}
f_{q,l\ P} \ \ \simeq \ \ 
\underbrace{\ 2^{\frac{1}{4}}\ G_F\,\!^\frac{1}{2}\ m_{q,l}\ }_{
\hbox{\normalsize  $ 4\ 10^{-6}\ m_{q,l}(\rm{MeV})$}
}\ 
\times\ \
\left\{\
\ba{ccl}
\, r\,x \!&=&\!\cos\zeta\ \cot\beta \ \ \ \ \ \ \ \hbox{ for \ $u,\, c,\, t$ \ quarks,} \vspace{2mm}\\
r/x \!&=& \!\cos\zeta \,\tan\beta \ \ \ \hbox{ for \ $d,\, s,\, b \ $ quarks \,and \,$e,\, \mu,\, \tau$ \ leptons.}
\ea\right.
\ee

\vspace{1mm}
These couplings may also be determined from the spin-1/spin-0 equivalence 
\cite{fayetu}, from the axial couplings of the $U$ when the extra-$U(1)$ symmetry is realized locally.
The $U$ current is obtained from the initial extra-$U(1)$ current, with an additional contribution proportional to  $\,J_Z= J_3-\sin^2\theta\,J_{\rm em}$  \,originating from $\,Z-U$ mixing effects, typically induced by $\,v_1\,$ and $\,v_2\,$ 
\,when $\,\tan\beta\neq 1\,$
\cite{fayetu,melange,fayetextrau}. 
This leads to the axial couplings of the $U$ to quarks and leptons \footnote{\small The vector couplings of the $U$  
are usually expressed as a linear combination  of the $B$ and $L$
(or $\,B-L$) and electromagnetic currents \cite{fayetextrau}, and may contribute to invisible meson decays, such as those of the 
$\,\pi^\circ, \, \eta,\, \eta', \, \psi$ or $\,\Upsilon$ \cite{fayetldmu,bes,besIII}. \,(Note that
$\psi\,$ and $\,\Upsilon\,$ cannot decay invisibly into dark matter particles, according to $\,\psi\,(\Upsilon)\to\chi\chi$,  
\,through the virtual production of a spin-0 boson \cite{fayetldmu}).
},

\vspace{-4mm}
\be
\label{fAtb}
f_{q,l\ A}\ \simeq \ 
\underbrace{\ 2^{-\frac{3}{4}}\ \ G_F\,\!^\frac{1}{2}\ \ m_{q,l}\ }_{
\hbox{\normalsize  $ 2\ 10^{-6}\ m_{U}(\rm{MeV})$}
}
\ \ \times\ \ 
\left\{\
\ba{ccllll}
\, r\,x \!&=&\!\cos\zeta\ \cot\beta \ \ \ \ \ \ \ \hbox{ for \ $u,\, c,\, t$ \ quarks,} \vspace{2mm}\\
r/x \!&=& \!\cos\zeta \,\tan\beta \ \ \ \hbox{ for \ $d,\, s,\, b \ $ quarks \,and \,$e,\, \mu,\, \tau$ \ leptons.}
\ea\right.
\ee
Using (\ref{effps}) we recover in this way 
the effective pseudoscalar couplings (\ref{couplage2})  of the $U$,
the same as for a standard axion or pseudoscalar $A$ in the MSSM, multiplied by the invisibility factor 
$\,r=\cos\zeta\,$.

\vspace{2mm}

\section{\boldmath New constraints \,from \,$\Upsilon\to \gamma$ + {\em \,invisible neutral}, \,and their consequences}

\begin{figure}[htb]
$$
\epsfig{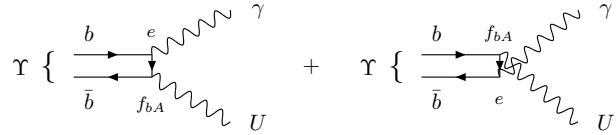}
$$
\caption{\small \em $\,\Upsilon\to\,\gamma\,U$ \,decay induced by the axial coupling $\,f_{bA}\,$. 
For a light $U$ the total amplitude is essentially the same as for a spin-0  $\,a$ 
\,with pseudoscalar coupling
$\,f_{bP}=f_{bA}\ \frac{2\,m_b}{m_U}$.}
\label{fig:upsilon}
\end{figure}

\begin{figure}[htb]
$$
\epsfig{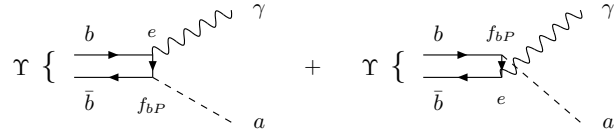}
$$
\caption{\small \em \,Production of a spin-0 pseudoscalar in $\,\Upsilon\,\to\,\gamma\,a\,$.}
\label{fig:ups2}
\end{figure}

\vspace{3mm}

The quarkonium branching ratios, obtained using (\ref{fAtb}) or (\ref{couplage2})
which introduce $\,r=\cos\zeta\,$,
\,may be expressed from the ratios of the pseudoscalar (or effective pseudoscalar) couplings 
to the quarks $\,f_{qP}\,$ to the elementary charge $e$ \cite{wilczekaxion,fayetu}.
They are given at lowest order, disregarding $m_U$ or $m_a$, \,by 
\be
\label{rate0}
\frac{B (\hbox{onium} \to \gamma \  U/a)}{B(\hbox{onium} \to \mu^+\mu^-)} \ \ =\ 
\ \frac{2\ f_{qP}^{\ 2}}{e^2}\ =\ 
\frac{G_F\,m_q^2}{\sqrt 2 \,\pi\,\alpha}\ \ (\,r^2 x^2\ \ \hbox{\small or}\ \ \frac{r^2}{x^2}\,)\ \ .
\ee

\vspace{1mm}
\noindent
The branching ratios, given by
\be
\label{rate1}
\left\{\ \ \ba{ccccc}
\displaystyle
\frac{B (\psi \to \gamma \  U/a)}{B(\psi \to \mu^+\mu^-)} &=&
\displaystyle
\frac{G_F\,m_c^2}{\sqrt 2 \,\pi\alpha}\ \ r^2\,x^2\ \ C_\psi\, F_\psi
&\simeq&  \displaystyle
8\ 10^{-4}\ \ r^2\,x^2\ \ C_\psi\, F_\psi\ \ ,
\vspace{3.5mm}\\ 
\displaystyle
\frac{B (\Upsilon \to \gamma \  U/a)}{B(\Upsilon \to \mu^+\mu^-)} &=& \displaystyle
\frac{G_F\,m_b^2}{\sqrt 2 \,\pi\alpha}\ \ \frac{r^2}{x^2}\ \ C_\Upsilon\,F_\Upsilon
&\simeq& \displaystyle
8\ 10^{-3}\ \ \frac{r^2}{x^2}\ \ \ C_\Upsilon\,F_\Upsilon\ \ ,
\ea \right.
\ee
and expressed in terms of $\ \cos\zeta \,\cot\beta\,= r\, x\ $ and
$\,\cos\zeta \,\tan\beta\,= r/x\,$,
\,are approximately equal to \cite{fayetu,fayetmezard,melange}:
\be
\label{rates}
\left\{ \  \  \begin{array}{ccl}
B \ (\ \psi \ \to \ \gamma \  \ U/a\ ) &\simeq &
\ \ 5 \ \ 10^{-5} \ \ \,\cos^2\zeta\,\cot^2\beta  \ \ \    C_{\psi} \,F_\psi\ \ ,
\nonumber 
\vspace{2.5mm}\cr
B\  (\ \Upsilon \ \to \ \gamma\  \ U/a\ ) &\simeq &
\  \ 2 \ \ 10^{-4} \ \ \,\cos^2\zeta\,\tan^2\beta \ \ \  C_{\Upsilon}   \,F_\Upsilon\ \ .
\cr
\end{array}   \right.
\ee

\noindent
$C_{\psi}$ and $C_{\Upsilon}$ take into account QCD radiative and relativistic corrections \cite{nason}, 
and are usually expected to be larger than $1/2$. 
$F$ is a phase space factor, equal to 
$\,1-\frac{m_U^{\,4}}{m_{\psi/\Upsilon}^{\,4}}\,$ \,or 
$ \,1-\frac{m_a^{\,2}}{m_{\psi/\Upsilon}^{\,2}}\,$,
\,with $\, 1-\frac{m_{U/a}^{\, 2}}{m_{\psi/\Upsilon}^{\,2}}= 2\,E_\gamma/m_{\hbox{\tiny$ \psi/\Upsilon$}}\,$.

\vspace{3mm}

\vbox{
\noindent
\underline{\it $\psi\,$ decays\,:}

\vspace{1mm}
For a $\,U$ with  mostly invisible decays (into $\,\nu\bar \nu\,$ or light dark matter particles), we get, from
$\, B\, (\psi \to   \gamma +\,  \hbox{invisible})   < 1.4 \  10^{-5}$,
$\ rx=\cos\zeta\,\cot\beta <  .75\,$ \cite{melange,fayetldmu,edwards},
\,which requires $|f_{cA}| < 1.5\ 10^{-6}\ m_U\hbox{(MeV)}\,$ for a pseudovector coupling, \,or $\,|f_{cP}| < 5\ 10^{-3}$
for the pseudoscalar coupling of a massless or light spin-0 boson $\,a\,$
that would not decay visibly within the detector. Very much the same limit, but taken as $\,|f_{cS}| <  10^{-2}$
as QCD corrections may now be somewhat larger \cite{nason}, \,applies, 
in a similar way \cite{wilczekhiggs}, to the coupling of a massless or light scalar boson with invisible decays.
If the invisible branching ratio $\,B_{\rm inv}$ (within the detector) is not $\,\simeq 1$, 
\,the limits get divided by $\,\sqrt {B_{\rm inv}}$, \,so that
\be
\label{limx}
rx=\cos\zeta\,\cot\beta \, <\,  .75/\sqrt {B_{\rm inv}}\ \ \ \Longleftrightarrow\ \ |f_{cA}| < 1.5\ 10^{-6}\ m_U\hbox{(MeV)}/\sqrt {B_{\rm inv}}\ , \ \ \hbox{or}\ \ |f_{cP}| < 5\ 10^{-3}/\sqrt {B_{\rm inv}}\ \ .
\ee
}

\vspace{3mm}
\noindent
\underline{\it $\Upsilon\,$ decays, \,and consequences for the $\,\psi\,$:}

\vspace{1mm}

The experimental limit on $\,\Upsilon \to \gamma +  \hbox{\it invisible}\,$ \cite{axionex} 
got improved with  the $\Upsilon(3S)$ \,by more than 4 by the BABAR coll. \cite{babar},
with a preliminary upper limit increasing from 3.2 to $3.5\ 10^{-6}$ 
when the mass of the unobserved neutral grows from 0 to 1 GeV, down to $\,.7\ 10^{-6}\,$ for 3 GeV,
and less than $4\ 10^{-6}$ for any mass up to 6 GeV.
We then get the new limits

\vspace{-5mm}
\be
\label{lim/x}
r/x=\cos\zeta\,\tan\beta \, <\, .2/\sqrt {B_{\rm inv}}\ \ \ \Longleftrightarrow\ \ |f_{bA}| < 4\ 10^{-7}\ m_U\hbox{(MeV)}/\sqrt {B_{\rm inv}}\ ,\ \ \hbox{or}\ \ |f_{bP}| < 4\ 10^{-3}/\sqrt {B_{\rm inv}}\ \ ,
\ee
which take into account the invisible branching 
ratio of the new boson. \,This remains valid for a new particle mass of up to about 5 GeV 
(taking into account the phase space factor $F_\Upsilon$), as long as invisible decay modes are present.
The limit (\ref{lim/x}) on the pseudoscalar (or effective pseudoscalar) coupling $\,f_{bP}$
is 5 times smaller than the standard Higgs coupling to $\,b$, $\,m_b/v \,\simeq \,2\ 10^{-2}$,
\,for an invisibly decaying boson.
It may also be applied to a scalar coupling $\,f_{bS}\,$ provided it is slightly relaxed, to \,$\ |f_{bS}| < \,6 \ 10^{-3}/\sqrt {B_{\rm inv}}\ $ \footnote{\small Considering that 
in the scalar case the correction factor $\,C_\Upsilon$ should be larger than $\ \simeq .2\,$, \,instead of .5,
\,leads to relax the bound by $\,\simeq 1.6\,$.}.

\vspace{3mm}
From
$\ r^2= \cos^2\zeta < .6\ \tan^2\beta\,/B_{\rm inv}\,$,
\,and $\ < 4\ 10^{-2}\ \cot^2\beta\,/B_{\rm inv}\,$,
\,we get the upper limit independent of $\,\beta$ on the invisibility parameter,

\vspace{-9mm}
\be
\label{limr}
r^2\ =\ \cos^2\zeta\ <\ .15\,/B_{\rm inv}\ ,
\ee
so that $\,a\,$ should be {\it \,mostly singlet} \ ($\,>85$\,\%),\ 
rather than doublet \ ($\,< 15$\,\%), for invisible decays of the new boson.
The $\,\Upsilon\,$ limit, expressed as a constraint on the doublet fraction
\be
\label{tan4}
\hbox{\it doublet fraction:}\ \ \ r^2=\cos^2\zeta\ <\ 4\,\%\ /\,(\tan^2\beta\,B_{\rm inv})\ \ ,
\ee
is stronger than the
$\,\psi\,$ one for $\tan\beta\,$ larger than $\simeq .5$. It 
requires that $a$ should be
\,($\,< 4$\,\%\, doublet, $\,>96$\,\%\, singlet) \,for $\,\tan\beta > 1$;
\,and $\,< .5$\,\%\, doublet \,for $\,\tan \beta > 3$, \,for invisible decays 
of the new boson.

\vspace{3mm}

The dependence on $B_{\rm \,inv}\,$ disappears when we evaluate the upper limit for the production,
in radiative decays of the $\,\psi$, \,of a new boson decaying invisibly.
The non-observation of a signal in $\,\Upsilon \to\gamma + \hbox{\it invisible neutral\,} \,$ 
decays implies a rather small branching ratio for the similar decay of the $\,\psi$,
\be
\label{limpsi}
B\ (\,\psi\to\gamma \ +\ \hbox{\it neutral}\,)\ \ B_{\rm inv}\ \lsim\ \, \ 10^{-6}/\tan^4\beta\ \ ,
\ee
i.e. $\,\lsim \,10^{-8}\,$ for $\,\tan\beta \,\gsim \,3\,$,
{\em \,independently of the invisible branching ratio $\,B_{\rm \,inv}\,$} \,(a result also applicable, 
with little change, to the production of a scalar particle).

\vspace{3mm}
\noindent
\underline{\it Constraints from $\,\Upsilon\,$ decays \,on the couplings \,to \,electrons\,:}

\vspace{1mm}

$\Upsilon\,$ results also have implications on the couplings of the new 
spin-1 or spin-0 boson to the {\it \,electron}, {\it \,muon\,} or $\,\tau$\, lepton. 
Eq.~(\ref{fAtb}) implies universality properties for the axial couplings of the $U$,
family-independent and identical for all charged leptons and down quarks \cite{melange}. This is also a consequence 
of the gauge invariance of the Yukawa couplings
responsible for their masses in a 2-Higgs-doublet model \cite{fayetextrau}, leading to
$\,f_{eA}\!=f_{\mu A}\!=f_{\tau A}=f_{dA}\!=f_{sA}\!=f_{bA}\,$.
\,It also reflects that the corresponding couplings of the pseudo\-scalar $a$ to down-quarks and charged leptons are proportional to their masses
(as expressed by (\ref{couplage2})), so that 

\vspace{-7mm}
\be
f_{eP}\ =\ f_{bP}\ \,m_e/m_b\ \ .
\ee
The new strong limit (\ref{lim/x}) 
on $\,f_{bA}$ then applies also to $\,f_{eA}$, \,severely restricting it to 

\vspace{-4mm}
\be
\label{feaups}
|f_{eA}|\  <\ 4\ \,10^{-7}\ \,m_U\hbox{(MeV)}\ /\sqrt{B_{\rm inv}}\ ,\ \ \ \ \hbox{or}\ \ \ \ 
|f_{eP}|\ <\ 4\ \,10^{-7}\, /\sqrt{B_{\rm inv}}\ \ .
\ee
The last limit on a pseudoscalar coupling $\,f_{eP}\,$
is {\it \,5 times smaller\,} than the standard Higgs coupling to the electron, $\,m_e/v\simeq 2\ 10^{-6}$,
\,for invisible decays of the new boson. 
As scalar couplings are also proportional to masses, 
with $\,h_1$ alone responsible for charged-lepton and down-quark masses so that 

\vspace{-4mm}
\be
f_{eS}\ =\ f_{bS}\ \,m_e/m_b\ \ ,
\ee
the last limit on $|f_{eP}|\,$ may also be applied to a {\it \,scalar\,} coupling provided it is slightly relaxed, as for
$\,|f_{bP}|\,$,
\,the effect of radiative corrections in $\,\Upsilon$ decays being larger in this case, so that
\be
\label{limfes}
|f_{eS}|\  < \  6\ \,10^{-7}\, /\sqrt{B_{\rm inv}}\ \ .
\ee
For a spin-1 $\,U$ boson the strong limit (\ref{feaups}) on $\,f_{eA}\,$  is in agreement with the results of parity-violation experiments in atomic physics, which imply 
a strong limit on $\,|f_{eA}\,f_{qV}|\,$
\cite{bouchiatfayet}.
It has implications on the size of the $\, e^+e^-\!\to\gamma \, U$ annihilation cross section, roughly proportional 
to $\,f_{eV}^2+f_{eA}^2$, \,which should then be very small for a light $\,U$, 
\,unless its vector coupling to the electron is significantly larger than the axial one.

\section{\boldmath Comparison with constraints from \boldmath
\ $\Upsilon\,\to \,\gamma + \,(\hbox{\bf \normalsize \em neutral} \to$ $ \mu^+\mu^-)\ $ decays}

The analysis applies as well to 
a relatively light spin-1 $U$ boson or spin-0 pseudoscalar $a$, or scalar, 
decaying visibly for example into $\,\mu^+\mu^-$ with a branching ratio $\,B_{\mu\mu}$. \,This was searched for recently
by the CLEO \cite{cleo} and BABAR \cite{babar2} collaborations with spin-0 particles in mind, but the results may be used
to constrain light spin-1 $\,U\,$ bosons as well, given the quasi-equivalence between their production rates 
in radiative $\,\Upsilon$ decays, \,pointed out long ago 
\cite{fayetu0,fayetu,fayetmezard} \,with the pseudoscalar $\ a\,$ already a mixing of doublet 
\,(``active'')\, and singlet \,(``inert'')\, components.
This is particularly relevant as many theoretical constructions now appeal
to such light weakly-coupled neutral bosons.

\vspace{2mm}

With 
$\,B\,(\Upsilon \to\gamma + \hbox{\it neutral\,})\ B_{\mu\mu}\,$ taken to be $\,\lsim \,2\ 10^{-6}\,$
in most of the mass range considered
(and always less than $\,5\ 10^{-6}\,$ at 90\,\% c.l.~\,up to nearly \,9 GeV, excepted for two small regions around the $\psi\,$ and $\psi'$), as compared to
$\,B\,(\Upsilon\to\gamma + \hbox{\it neutral\,})\ B_{\rm inv}\, \lsim \,3.5 \ 10^{-6}$,
\,we can rescale (\ref{lim/x}) into $\ r/x=\cos\zeta\,\tan\beta \, \lsim\, .15/\sqrt {B_{\mu\mu}}\,$, \,so that
\be
\label{lim/x2}
|f_{bA}|\, \lsim \, 3\ 10^{-7}\ m_U\hbox{(MeV)}/\sqrt {B_{\mu\mu}}\ ,\ \ \ |f_{bP}|\,  \lsim \, 3\ 10^{-3}/\sqrt {B_{\mu\mu}}\ ,
\ \ \hbox{or}\ \ 
|f_{bS}|\,  \lsim\,  5\ 10^{-3}/\sqrt {B_{\mu\mu}}\ \ .
\ee
With a more conservative experimental upper limit $\,\lsim 4\ 10^{-6}$, \,the above limits should be slightly relaxed, to

\vspace{-4mm}
\be
\label{lim/x3}
\ba{l}
r/x=\cos\zeta\,\tan\beta \, \lsim\, .2/\sqrt {B_{\mu\mu}}\ \ \ \ \ \ \ \Longleftrightarrow\ \ 
\vspace{4mm}\\
\hspace{2cm} |f_{bA}| \lsim 4\ 10^{-7}\ m_U\hbox{(MeV)}/\sqrt {B_{\mu\mu}}\ ,\ \ \ \ |f_{bP}| \lsim 4\ 10^{-3}/\sqrt {B_{\mu\mu}}\ ,\ \  \hbox{or}\ \ |f_{bS}|\,  \lsim\,  6\ 10^{-3}/\sqrt {B_{\mu\mu}}\ \ .
\ea
\ee

\vspace{3mm}

These limits may or may not be more constraining that (\ref{lim/x}), depending on whether or not $\,B_{\mu\mu}$ is larger than 
$\,\approx B_{\rm inv\,}$.
\,As an illustrative example a 1 GeV  $\,U\,$ boson could have $\,B_{\rm \,inv}\! \approx 16\,\%  \,$ and 
$\,B_{\mu\mu}\!\approx 10\,\% \,$ \cite{fayetu}, \,if we ignore the possibility of light dark matter particles \cite{ldm} which could make $\,B_{\rm \,inv}$ very close to 1.
See also  \cite{domingo} for a discussion of the upper limit on $\,\cos\zeta\,\tan\beta\ $ 
for a spin-0 pseudoscalar $\,a\,$ in the NMSSM, using CLEO results \cite{cleo}.

\vspace{2mm}

Eq.~(\ref{lim/x2}) translates into
\be
\label{tan42}
\hbox{\it doublet fraction:}\ \ \ r^2=\cos^2\zeta\ \lsim \ 2\,\%\ /\,(\tan^2\beta\ B_{\mu\mu})\ \ .
\ee
The dependence on $B_{\mu\mu}$ disappears when we evaluate the upper limit for the production,
in radiative decays of the $\,\psi$, \,of a new spin-1 or spin-0 boson decaying into $\,\mu^+\mu^-$.
The non-observation of a signal in $\,\Upsilon \to\gamma+ (\hbox{\it neutral\,} \to\,\mu^+\mu^-)$ 
decays thus also implies a small branching ratio in the similar decay of the $\,\psi$
(very much as we saw in (\ref{limpsi}) for the invisible decays):
\be
\label{limpsi2}
B\ (\,\psi\,\to\,\gamma +\,\hbox{\it neutral\,}\,)\ \ B_{\mu\mu}\ \lsim\ \, 5\ \,10^{-7}/\tan^4\beta\ \ ,
\ee
i.e. $\,\lsim \,5 \ 10^{-9}$ for $\,\tan\beta\, \gsim \,3\,$, {\em \,independently of $\,B_{\mu\mu}\,$},
\,a result also applicable to the production of a scalar particle.
Again limits on $\,b$ couplings  may be translated into limits on the pseudovector, pseudoscalar or scalar couplings 
to the electron, leading, very much as in (\ref{feaups},\ref{limfes}), to
\be
\label{feaups2}
|f_{eA}|\  \lsim\ 3\ \,10^{-7}\ \,m_U\hbox{(MeV)}\ /\sqrt{B_{\mu\mu}}\ \,, \ \ \ \ \ \ \ 
|f_{eP}|\ \lsim \ 3\ \,10^{-7}\, /\sqrt{B_{\mu\mu}}\ \ \ \ \ \hbox{or}\ \ \ \ 
|f_{eS}|\ \lsim \ 5\ \,10^{-7}\, /\sqrt{B_{\mu\mu}}\ \ .
\ee

\section{Conclusions}

Theories with 2 Higgs doublets may allow for a broken extra-$U(1)$ symmetry,
local or global, acting axially on quarks and leptons, and leading to new light neutral spin-1 or spin-0 bosons.
This occurs naturally in supersymmetric extensions of the standard model  
with a trilinear $\,\lambda \, H_1 H_2 \,S\,$ superpotential.
The extra $U(1)$ may be gauged, as in the USSM, and the new spin-1 boson $U$, 
which eliminates an axionlike pseudoscalar
$a$, could be light if the corresponding gauge coupling is small.
This may have a more profound origin with a possible connection of the extra-$U(1)$ and $\,U$ boson 
with the gravitino and gravity itself, which would not be so surprising as the pseudoscalar $\,a$, \,and associated scalar partner, interact proportionally to masses.

\vspace{1mm}
$\Upsilon$ decays constrain
an axial, pseudoscalar or scalar coupling to the $\,b\,$ to  
$\,f_{bA}<4\ 10^{-7}\ m_U$ (MeV)$/\sqrt{B_{\hbox{\scriptsize inv}}}\,$,
\,$f_{bP}< 4\ 10^{-3}/\sqrt{B_{\hbox{\scriptsize inv}}}\ $ (\,5 times less than the standard Higgs coupling, 
for invisible decays of the new boson) 
or $\,f_{bS}< 6\ 10^{-3}/\sqrt{B_{\hbox{\scriptsize inv}}}\,$, \,respectively,
also constraining strongly their couplings to the electron
(to e.g. $\!< 4\ 10^{-7}/\sqrt{B_{\rm inv}}\,$ for a pseudo\-scalar).
\,Similar limits have been obtained from searches for 
$\ \Upsilon\to\gamma+(\hbox{\em neutral}\to \mu^+\mu^-)$, \,for $\,m_{\hbox{\footnotesize neut.}}>2\,m_\mu$,
\,which, altogether, strongly constrain the rates for both $\, \psi\to\gamma+\hbox{\em invisible neutral}\ $ 
and $\ \psi\to\gamma+(\hbox{\em neutral}\to \mu^+\mu^-)$.

\vspace{1mm}
The results apply, generically, in a large class of theoretical models
involving an extra-$U(1)$ symmetry, 
either global or local, and whether supersymmetry is present or not, even if 2-higgs doublet susy extensions of the 
standard model (N/nMSSM, USSM, ...~) provide the most natural framework and best motivation.
They are relevant for a variety of experimental searches, including quarkonium decays
(e.g.~at BES III or with a super $B$ factory \cite{besIII,factory})  and other experiments 
relying on the coupling of the new neutral boson 
to the electron, muon or $\,\tau$ lepton.

\vspace{1mm}

The search for light weakly coupled particles such as goldstinos/gravitinos,
$U$ bosons, axions, axionlike or dilatonlike particles, ...\,, constitutes
a direction to be further explored,
in complement of the high-energy frontier 
at the Tevatron, LHC, and ILC. This may also contribute to the understanding of high-energy physics,
with the very weak couplings of such light particles in close relation with the mass spectrum.

\end{document}